# Audio-JEPA: Joint-Embedding Predictive Architecture for Audio Representation Learning

Ludovic TUNCAY [1] ludovic.tuncay@irit.fr
Etienne LABBE [1] etienne.labbe@irit.fr
Emmanouil BENETOS [2] emmanouil.benetos@qmul.ac.uk
Thomas PELLEGRINI [1] thomas.pellegrini@irit.fr

[1] IRIT, Université de Toulouse, CNRS, Toulouse INP, Toulouse, France
[2] School of Electronic Engineering and Computer Science, Queen Mary University of London, UK

***Abstract*—Building on the Joint-Embedding Predictive Architecture (JEPA) paradigm, a recent self-supervised learning framework that predicts latent representations of masked regions in high-level feature spaces, we propose Audio-JEPA (Audio Joint-Embedding Predictive Architecture), tailored specifically for audio data. Audio-JEPA uses a simple Vision Transformer backbone to predict latent representations of masked spectrogram patches rather than reconstructing raw audio. We pre-train on unlabeled AudioSet clips (10s, 32kHz) with random patch masking on mel-spectrograms. We evaluate on the X-ARES suite covering speech, music, and environmental-sound tasks. Although our implementation is a straightforward translation of the original model to audio, the results still show comparable performance to wav2vec 2.0 and data2vec while using less than one-fifth of their training data and with no hyper-parameter tuning. All code and pretrained checkpoints are available on GitHub[1].**

## I. Introduction

Self-Supervised Learning (SSL) has revolutionized representation learning for speech and audio, enabling models to learn from unlabeled data and excel in diverse downstream tasks [1, 2, 3, 4]. Early SSL approaches for audio, such as contrastive predictive coding and wav2vec 2.0, learned latent speech representations by masking the input and solving a contrastive task over latent codes [5]. Follow-up methods like HuBERT [1] introduced offline clustering to generate pseudo-labels for masked audio segments and WavLM [6] applied data augmentation and denoising to improve robustness in speech representation learning. More recently, latent prediction approaches have gained traction: data2vec [7] and its efficient successor data2vec 2.0 [8] employ a teacher–student framework to predict contextualized latent representations of the input, achieving strong results across vision, speech, and language tasks. In the audio domain, Niizumi et al. introduced Masked Modeling Duo (M2D) [4], which uses two networks (online and momentum encoder) to predict masked patch embeddings and attained state-of-the-art results on numerous audio benchmarks.

In computer vision, a new paradigm called Joint-Embedding Predictive Architecture (JEPA) [9, 10, 11] has been proposed to predict hidden content in a high-level latent space instead of pixel space. Notably, the image-based I-JEPA [10] model demonstrated that predicting representations of masked image regions can yield powerful visual features. The JEPA approach differs from prior masked reconstruction methods by focusing on semantic latent prediction rather than lower-level signal reconstruction. Inspired by these advances, an audio version of JEPA (termed A-JEPA) was recently described by Fei et al [12]. Their A-JEPA encodes a spectrogram "context" part and predicts the latent representations of masked "target" regions using a momentum-updated target encoder. During pre-training, they anneal the mask from fully random toward a spec-augment-style structured scheme. In contrast, our Audio-JEPA sticks with purely random masking throughout for simplicity and maximal generality. At the time of writing, no official implementation or checkpoints for A-JEPA are available, motivating our from-scratch development. In the musical domain, Stem-JEPA [13] adapts the JEPA paradigm to multi-track recordings by jointly training an encoder and predictor to forecast embeddings of compatible instrument stems. While utilizing the JEPA backbone, Stem-JEPA differs methodologically as it masks entire instrument stems instead of individual spectrogram patches.

Our work targets the ICME 2025 Audio Encoder Capability Challenge, where the goal is to learn general audio representations that perform well across a broad suite of tasks. We present our Audio-JEPA implementation - developed from scratch following the I-JEPA paradigm - and benchmark it on the challenge's eXtensive Audio Representation and Evaluation Suite (X-ARES). Our contributions include: (1) adapting the JEPA masked latent prediction framework to audio spectrogram inputs using a Vision Transformer (ViT) backbone; (2) an extensive evaluation against prior self-supervised audio models on standard downstream tasks, assessing Audio-JEPA via both linear probing and k-nearest-neighbor evaluation.

## II. Related Work

### A. Self-Supervised Audio Representation Learning

Early work in SSL for audio focused on predicting future or missing parts of the waveform. wav2vec 2.0 [5] pioneered masking in the latent speech representations and training the model to identify the true quantized latent of a masked segment among distractors (a contrastive loss). This approach enabled models to learn rich speech features and achieved remarkable results on speech recognition with limited labeled data. Building on this, HuBERT (Hidden-Unit BERT) [1] introduced a BERT-like masked prediction where the model predicts cluster assignments of masked audio frames.

[1] https://github.com/LudovicTuncay/Audio-JEPA

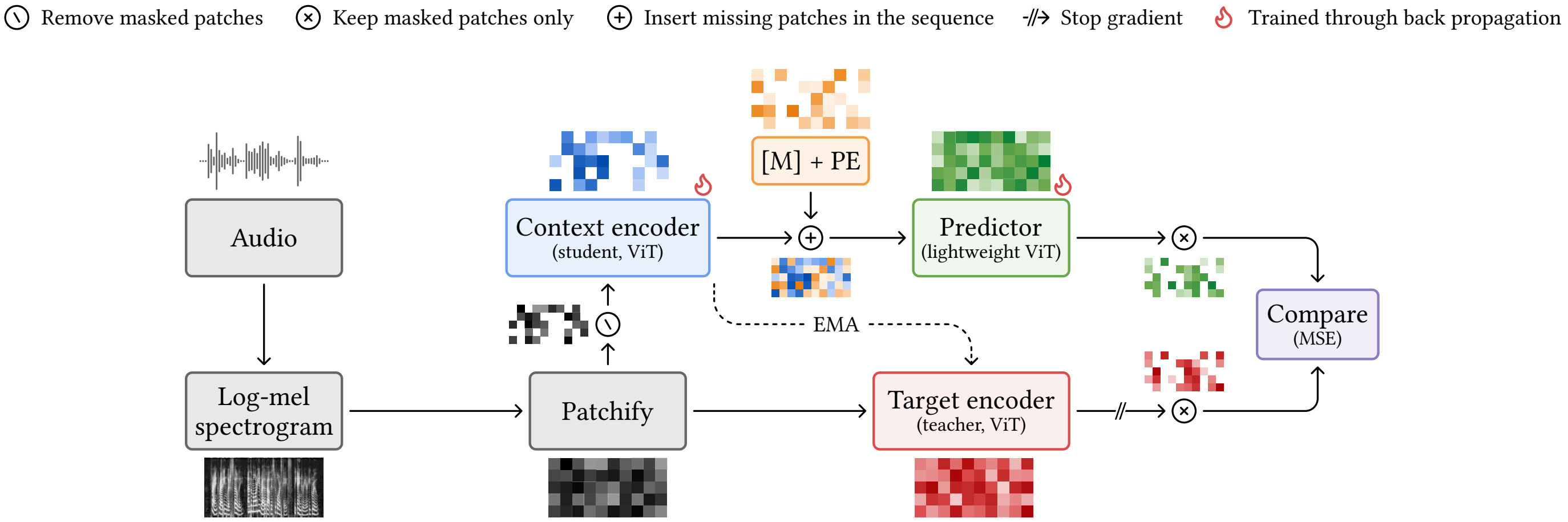

Fig. 1. **Audio-JEPA architecture**. Mel-spectrogram patches are split into visible and masked sets. A context encoder embeds visible patches, a lightweight predictor reconstructs masked-patch embeddings, and a momentum-updated target encoder provides targets. Training minimizes the mean squared Euclidean distance between predicted and target embeddings. $[M] + PE$ denotes the learned mask token with the positional encoding of the corresponding masked patch added.

HuBERT uses an offline k-means on acoustic features to provide target labels, and by iteratively refining these labels, it learns high-level speech units, matching wav2vec 2.0 performance on ASR benchmarks. Facebook's data2vec presented a modality-general SSL approach: instead of contrastive or classification targets, data2vec [7] trains a student network to regress the contextualized embeddings produced by a teacher network (an Exponential Moving Average (EMA) of the student) for masked portions of the input. data2vec 2.0 [8] improved the efficiency of this method by not encoding masked tokens and using a lightweight decoder, achieving similar accuracies to Masked Autoencoders [14] in a fraction of the training time and matching wav2vec 2.0 on speech tasks with over a 10× speedup. These latent regression approaches eliminate the need for discrete targets and have set strong baselines in audio.

### B. *Masked Prediction with Dual Networks*

The use of two networks (online/target) for masked prediction has also been explored in specialized audio SSL methods. M2D (Masked Modeling Duo) [4] employs an online network that sees the unmasked patches and a momentum target network that encodes only the masked patches. The online network predicts the target network's representation of the masked region, encouraging both networks to effectively model the input. This design, inspired by Masked Autoencoders but working in representation space, led M2D to state-of-the-art results on a range of audio classification tasks (environmental sound, speaker ID, music genre, etc.). Notably, M2D achieved top performance on datasets like UrbanSound8K [15], VoxCeleb1 [16], GTZAN [17], and SpeechCommands [18] with a single universal model. Such results highlight the power of using latent prediction instead of raw signal reconstruction for learning transferable audio features. Other contemporary models include WavLM [6], which extended HuBERT with simulated noisy inputs and achieved strong results on both speech recognition and classification tasks.

### C. *Joint-Embedding Predictive Architectures*

Rather than predicting low-level details of masked inputs, JEPA methods aim to predict higher-level representations. The image-based I-JEPA [10] demonstrated that a ViT can learn excellent representations by predicting the latent representations of masked image patches, as opposed to generating pixels. By operating in the feature space, I-JEPA forces the model to capture abstract semantic information and ignore minute pixel-level differences. The concept has since been extended to other modalities and combinations (e.g., TI-JEPA [19] for text–image, GeoJEPA [20] for geospatial data, etc.), showing JEPA's flexibility. For audio, Fei et al. recently proposed A-JEPA [12], applying the same principle to spectrogram inputs. While A-JEPA and M2D both adopt a dual-network masked-prediction framework, they differ in how the target encoder is applied: M2D processes only the masked spectrogram patches, whereas JEPA processes the entire spectrogram (context + masked). This richer context enables more detailed representations of the masked regions. Their design uses a context encoder to process unmasked spectrogram patches and a target encoder (the EMA of the context network) to encode masked regions, with a lightweight predictor network aligning the two in latent space. Our work is directly inspired by this approach but contrary to M2D, JEPA does not require data augmentation and the whole spectrogram is seen by the target encoder. We evaluate Audio-JEPA in our experiments, underlining how it bridges the gap between vision-style masked modeling and audio understanding.

## III. Proposed Method: JEPA for Audio

In this section, we describe our adaptation of the JEPA paradigm to the audio domain, which we call Audio-JEPA. We first present the overall architecture, then detail the self-supervised training objectives, and finally highlight the audio-specific design choices that make A-JEPA effective on diverse sound data.

### A. Overall architecture

As shown in the Fig. 1, our Audio-JEPA model consists of three main modules:

1) **Context encoder:** Processes the "visible" subset of Mel-spectrogram patches
2) **Target Encoder:** Provides stable target embeddings via an Exponential Moving Average (EMA) of the context encoder's parameters
3) **Lightweight Predictor Network:** Takes context embeddings and predicts latent representations for masked ("target") patches

Upon converting an input waveform to a Mel-spectrogram and partitioning it into non-overlapping time–frequency patches, we randomly mask a fixed proportion of patches. The context encoder embeds the remaining visible patches, producing a context representation. The lightweight predictor network then reconstructs embeddings for the masked patches. In parallel, the target encoder (updated by EMA rather than gradient descent) encodes the true masked patches. Training minimizes the average L2 distance between the predictor's outputs and the target encoder's embeddings, with stop-gradient applied between the predictor and the target encoder. This implementation is a direct adaptation of I-JEPA to the audio domain, by considering the spectrogram as a single channel, possibly non-square, image.

### B. Training objective

We train Audio-JEPA using the average $L_2$ distance between the predicted patch-level representations and the target patch-level representation in the masked parts. Formally, let

$$c_i = f_{\text{ctx}}\left(x_{\setminus M}\right)_i, \quad \hat{c}_j = g_{\text{pred}}(c), \quad t_j = f_{\text{tgt}}(x)_j \tag{1}$$

where $x$ are the Mel-spectrogram patches, $x_{\setminus M}$ are the visible patches, $f_{\text{ctx}}$ and $f_{\text{tgt}}$ are the context and target encoders respectively, and $g_{\text{pred}}$ the lightweight predictor. The loss is then

$$\mathcal{L} = \frac{1}{|M|}\sum_{j \in M} \|\hat{c}_j - t_j\|_2^2 \tag{2}$$

We update $f_{\text{ctx}}$ and $g_{\text{pred}}$ parameters via backpropagation, while $f_{\text{tgt}}$ parameters are updated as

$$\theta_{\text{tgt}} \leftarrow \tau\theta_{\text{tgt}} + (1-\tau)\theta_{\text{ctx}} \tag{3}$$

with $\tau$ the EMA decay factor. This design stabilizes target representation and prevents collapse.

### C. Evaluation

Our assessment follows the eXtensive Audio Representation and Evaluation Suite (X-ARES)[2], which brings together 21 publicly available audio datasets spanning a variety of tasks and domains. Using the frozen target encoder in the evaluation, we employ two complementary evaluation strategies drawn from X-ARES:

a) *Linear Probing (MLP):* For each downstream task, we freeze the pre-trained encoder and attach a single linear layer. This classifier is trained on the task's labeled data using a fixed set of hyperparameters. By holding the original model weights constant, this procedure reveals how readily the learned representations can be linearly separated and adapted to new tasks.

b) *k-Nearest Neighbors (kNN):* Without any additional training, we directly apply a kNN classifier to the frozen embeddings. This non-parametric evaluation highlights the raw discriminative power of the representations. Although it may underperform more sophisticated fine-tuning methods, kNN offers a strict baseline for the intrinsic quality of the learned features.

Due to the architecture and loss, the model's outputs are not guaranteed to be linearly separable as explained in the V-JEPA paper [11]. Therefore, we do not expect good results in that section. However, we should observe decent performance in the kNN task.

## IV. Implementation Details

In this section we summarize the key implementation choices and hyperparameters used to train A-JEPA. Tables 2–3 collect the most important settings; for clarity we defer dataset splitting, and hardware details to Section V.

### A. Data processing

We work with 1921982 AudioSet clips resampled to 32kHz and 10s duration totaling to 5338 hours of audio. Each waveform is converted to a 128-band Mel-spectrogram with 256 time bins (via a frame size and hop chosen accordingly such that the frame size is 2.5 times the size of the hop). Per example, we randomly sample 40 %–60 % (exact value per batch is uniformly sampled) of the patches indices to be masked. Each batch contains 256 audio clips. Preliminary experiments showed the block masking strategy from I-JEPA yielded lower performance than random masking.

### B. Model Architecture

In Table I you can see the exact ViT hyperparameters used for each module. The context and target encoders share the same ViT configuration with 16×16 patches, a 768-dimensional embedding, 12 layers, 12 attention heads and an MLP ratio of 4.0. The target encoder is kept architecturally identical to the context encoder and updated via EMA where the parameter $\tau$ is set in the same way as in BYOL [21]. The predictor uses an embedding size of 384, a head count of 12 and contains 6 layers. MLP ratio remains the same at 4.0. The predictor re-projects the embeddings to 768 after going through the ViT so that its outputs can to be compared with those of the target encoder. The total number of trainable parameters during training is 96.7M, with 85.4M parameters used at inference since the predictor is not used.

### C. Optimization and Scheduling

We train using AdamW [22] ($\beta_1 = 0.9, \beta_2 = 0.95$, weight decay$= 0.05$) and an initial learning rate of $3 \cdot 10^{-4}$. A warmup-cosine scheduler ramps from $1 \cdot 10^{-6}$ to $3 \cdot 10^{-4}$ over 1,000 steps, then anneals to zero.

[2]https://github.com/jimbozhang/xares

TABLE I
Encoders and predictor architecture hyperparameters. The total number of trainable parameters during training is 96.7M, with 85.4M used at inference since the predictor is not needed.

| Hyperparameter | Context / Target Encoders | Predictor |
|---|---|---|
| Patch size | $16 \times 16$ | - |
| Embedding dim | 768 | 384 |
| Depth | 12 | 6 |
| Heads | 12 | 12 |
| MLP ratio | 4.0 | 4.0 |
| # Parameters | 85.4M (each) | 11.3M |

# V. Experimental Setup

We train on 4 NVIDIA V100 GPUs, with a total batch size of 256 clips. In other words, each batch contained somewhere close to 42.7 minutes of audio. Training took place for 100,000 steps (~13 epochs) which took 14 hours to complete. This training necessitated significantly less resources than Wav2Vec2 Base [5] and data2vec [7] where each trained for 400k steps with bigger batches of 1.6h and 63 minutes of audio respectively.

# VI. Results

## A. Linear-probe (MLP) performance

We first probe Audio-JEPA's representations with a small MLP head to assess their linear separability across tasks. Table II reports Audio-JEPA's performance on 20 of the 21 X-ARES datasets (all except LibriSpeech-100h [34]). The model reaches first or second place on several benchmarks but falls to last on roughly half of them. In particular, Audio-JEPA underperforms significantly on Fluent Speech Commands and Speech Commands V1. As noted in Section III.C.b, Audio-JEPA is at a disadvantage under linear-probe evaluation as Audio-JEPA's embedding space is not guaranteed to be linearly separable due to the training objective favoring embedding cohesion. As experimented in [11], using attentive pooling might help in that situation.

## B. kNN performance

We next evaluate pure embedding quality by k-nearest-neighbor classification on the 16 X-ARES tasks compatible with this probe. Table III shows that Audio-JEPA achieves first place on 3 datasets (ESC-50, FMA-small, GTZAN) and second place on 7 more, outperforming wav2vec 2.0 and data2vec despite using far less pre-training data and resources. Conversely, it ranks last on about a third of the tasks, underperforming substantially on those same datasets as when evaluated with linear-probing. Because kNN probes frozen embeddings directly and without extra classifier capacity, these results may better reflect the true representational power of Audio-JEPA across music, environmental sounds and speech domains.

TABLE II
Linear-probing evaluation results on the X-ARES evaluation suite, comparing our Audio-JEPA implementation with wav2vec 2.0 and data2vec. Scores are reported as given by X-ARES; **bold** indicates the best performance, and underline indicates the second-best.

| Dataset | Audio-JEPA (ours) | Wav2Vec2 [5] | Data2Vec [7] |
|---|---|---|---|
| asda | | asdas | fdgdfg |
| ASV2015 [23] | 0.898 | 0.924 | **0.937** |
| Clotho [24] | **0.014** | **0.014** | 0.008 |
| CREMA-D [25, 26] | 0.427 | **0.541** | 0.523 |
| DESED [27, 28] | 0.306 | **0.313** | 0.136 |
| ESC-50 [29] | 0.338 | **0.510** | 0.229 |
| Fluent Speech Command [18] | 0.025 | 0.468 | **0.978** |
| Free Music Archive small [30] | **0.553** | 0.469 | 0.334 |
| FSD18-Kaggle [31] | 0.212 | **0.241** | 0.153 |
| FSD50k [32] | 0.151 | **0.166** | 0.085 |
| GTZAN Genre [17] | 0.628 | **0.630** | 0.448 |
| LibriCount [33] | 0.471 | **0.583** | 0.492 |
| LibriSpeech-MF [34] | 0.883 | **0.948** | 0.752 |
| NSynth-Instruments [35] | 0.404 | **0.443** | 0.336 |
| RAVDESS [36] | 0.303 | 0.442 | **0.467** |
| Speech Commands V1 [18] | 0.152 | 0.714 | **0.927** |
| UrbanSound 8k [15] | 0.585 | **0.659** | 0.426 |
| Vocal Imitation [37] | 0.056 | **0.147** | 0.128 |
| VocalSound [38] | 0.526 | 0.768 | **0.803** |
| VoxCeleb1 [16] | 0.041 | **0.340** | 0.105 |
| VoxLingua33 [39] | 0.093 | **0.553** | **0.620** |

## C. Closing summary

Overall, Audio-JEPA demonstrates that a simple mask-prediction objective can yield high-quality audio embeddings with far less pre-training data. Under kNN evaluation, Audio-JEPA often matches or surpasses baselines on music and general-sound tasks, even though those baselines underwent far more extensive pre-training, confirming the strength of latent-space prediction. Linear-probe results expose the limitations of a single-layer head for Audio-JEPA, but also point to clear remedies (attentive pooling or small multi-layer probes). Across both evaluation methods, Audio-JEPA is weakest on tasks requiring fine-grained speech discrimination (e.g. speaker verification and keyword spotting), indicating that specialized data could improve these cases. These results validate Audio-JEPA as a data-efficient foundation for audio representation learning and set the stage for the architectural and tuning improvements detailed in the Conclusion.

TABLE III

kNN evaluation results on the X-ARES evaluation suite, directly comparing frozen embeddings from Audio-JEPA against wav2vec 2.0 and data2vec. Scores are reported as given by X-ARES; **bold** indicates the best performance, and underline indicates the second-best.

| Dataset | Audio-JEPA (ours) | Wav2Vec2 [5] | Data2Vec [7] |
|---|---|---|---|
| ASV2015 [23] | 0.927 | 0.858 | **0.942** |
| CREMA-D [25] | 0.267 | 0.221 | **0.351** |
| ESC-50 [29] | **0.140** | 0.081 | 0.040 |
| Fluent Speech Command [18] | 0.009 | 0.017 | **0.630** |
| Free Music Archive small [30] | **0.449** | 0.251 | 0.106 |
| GTZAN Genre [17] | **0.452** | 0.303 | 0.108 |
| LibriCount [33] | 0.307 | **0.311** | 0.176 |
| LibriSpeech-MF [34] | 0.545 | 0.606 | **0.724** |
| NSynth-Instruments [35] | 0.170 | **0.251** | 0.179 |
| RAVDESS [36] | 0.215 | 0.169 | **0.313** |
| Speech Commands V1 [18] | 0.044 | 0.208 | **0.852** |
| UrbanSound 8k [15] | 0.303 | **0.339** | 0.156 |
| Vocal Imitation [37] | 0.017 | 0.010 | **0.018** |
| VocalSound [38] | 0.256 | 0.269 | **0.308** |
| VoxCeleb1 [16] | 0.002 | 0.003 | **0.033** |
| VoxLingua33 [39] | 0.057 | 0.034 | **0.058** |

## VII. Conclusion

We introduced **Audio-JEPA**, the first open-source, from-scratch adaptation of the Joint-Embedding Predictive Architecture to audio. Pre-trained on AudioSet with off-the-shelf hyper-parameters, Audio-JEPA delivers competitive performance on the X-ARES. The method compares especially well under kNN evaluation, confirming that predicting masked latent targets, rather than the waveform or discrete labels, is a useful inductive bias for general-purpose audio representation learning.

Looking forward, we identify three straightforward upgrades:

1) **Attention-pooling head.** Replacing the single-frame MLP used in our linear-probe evaluation with a lightweight attention-pooling block, as proposed for V-JEPA [11], could yield fairer comparisons and narrow the current linear-probe gap.
2) **Modern backbones and positional encodings.** Swapping the vanilla ViT for recent audio transformers (e.g. ConvFormer [40] or CAFormer [41]) and testing rotary or conditional sine-cosine encodings should improve modelling of long-range temporal cues.
3) **Hyper-parameter tuning.** A systematic sweep of mask ratio, EMA decay and optimizer settings, currently untouched, is likely to uncover additional headroom.

By open-sourcing our code and checkpoints, we hope to establish Audio-JEPA as a solid starting point for the community to explore these directions and to further unify JEPA research across vision and now audio.

## Acknowledgement

This work was granted access to the HPC resources of IDRIS under the allocation AD011014754 made by GENCI. Support from the ANR-3IA Artificial and Natural Intelligence Toulouse Institute ANITI (ANR-19-PI3A-0004) is gratefully acknowledged.